\documentstyle{elsart}

\newcommand{\intercal}{T}
\newcommand{\text}{\rm}
\begin{document}
\begin{frontmatter}
\title { Symmetry-projected Hartree-Fock-Bogoliubov Equations}
\author[Munchen]{Javid A. Sheikh},
\author[Munchen]{Peter Ring}
\address[Munchen]{Physik-Department, Technische Universit\"at M\"unchen,
D-85747 Garching, Germany }

\begin{abstract}
Symmetry-projected Hartree-Fock-Bogoliubov (HFB) equations
are derived using the variational ansatz for the
generalized one-body density-matrix in the Valatin form. It
is shown that the projected-energy functional can be
completely expressed in terms of the HFB density-matrix and
the pairing-tensor.  The variation of this projected-energy
is shown to result in HFB equations with modified
expressions for the pairing-potential $\Delta$ and the 
Hartree-Fock field $\Gamma$.  The expressions for these
quantities are explicitly derived for the case of particle
number-projection.  The numerical applicability of this
projection method is studied in an exactly soluble model
of a deformed single-j shell.
\end{abstract}
\end{frontmatter}

%\pacs{PACS numbers : 21.60.Cs, 21.10.Hw, 21.10.Ky, 27.50.+e}

\section{Introduction}

The concept of spontaneous symmetry breaking in mean-field
theories has played a central role in our understanding of
many-body problems \cite{thou}. The exact solution of a
quantum mechanical many-body problem is impossible except
for a few-body system and it is therefore necessary to
adopt approximate methods. The most popular of these
methods has been mean-field theory based on effective
forces. The basic idea of the mean-field approach is to
approximate the ''unknown'' many-body wavefunction by a
Slater-determinant of single-particle or quasiparticle
configurations. In this way the many-body problem is
reduced to an effective one-body problem. In other words,
the system of interacting particles is transformed to a
system of non-interacting particles or quasiparticles. The
effective one-body potential is obtained from the many-body
Hamiltonian by using the Hartree-Fock (HF) or
Hartree-Fock-Bogoliubov (HFB) approaches \cite{RS.80}. The
HF method has first been introduced to solve the atomic
many-body problem and it was shown that the many-electron
problem can be reduced to one-electron problem with an
effective potential which is determined by the interactions
among all the electrons.  This method has been successfully
applied to the nuclear many-body problem, although at first
sight it appears conceptually contradictory to use a
mean-field concept to a strongly interacting system. It was
established later \cite{bethe71} that the Pauli-exclusion
principle favours the description of the nuclear many-body
problem in terms of the mean-field concept based on
effective interactions.

The HF approximation describes the long-range part,
referred to as the particle-hole ($p-h$) channel, of the
effective interaction and, indeed, is quite appropriate for
the atomic many-body problem. However, for the nuclear
many-body problem it can only be used in closed shell
configurations. For open shells, strong short-range
pairing-correlations in the particle-particle ($p-p$)
channel need to be considered. In oder to treat the $p-p$
channel, the Bardeen-Cooper-Schrieffer (BCS) approximation
\cite{bcs}, originally introduced to explain
superconductivity in metalls, has been applied to the
nuclear problem with remarkable success
\cite{BMP.58,Bel.59}. For a complete description, both
$p-p$ and $p-h$ channels need to be considered in a
self-consistent manner. This is achieved in the generalised
Hartree-Fock-Bogloliubov formalism
\cite{bogo}. 

It is an essential feature of mean-field theories that
they can violate symmetries, i.e. the approximate product
wavefunctions does not obey the same symmetries as the
underlying two-body Hamiltonian. The broken symmetries, for
instance, include  translational symmetry, rotational
symmetry, and the gauge symmetries connected with the
particle numbers. Since the exact solution of the many-body
Schr\"odinger equation obeys these symmetries being
eigen-function of a complete set of generators of the
corresponding symmetry group, the violation of symmetries
in the first place is a very undesirable feature. However,
the concept of symmetry violation allows to take into
account correlations in a very simplified form by employing 
product states which are 
connected with phase-transitions and provide a relatively
easy understanding of very complicated phenomena.
Nevertheless, in systems with finite particle number,
such as atomic nuclei these phase-transitions are never
sharp and are smeared out by fluctuations. It is one of
the disadvantages of the mean-field approximation that it
leads to sharp phase-transitions even in finite systems,
where the exact solution which takes into account
fluctuations beyond the mean-field shows only a
smooth transition between the two phases.

In order to have a better description of finite systems in
the vicinity of phase-transitions, it is necessary to
include fluctuations. One very powerfull way to consider such
fluctuations is through the restoration of the broken symmetries by
using the projection methods \cite{PY.57}.

There exists a vast amount of literature on such projection methods
at zero temperature (see, for instance, Ref. \cite{RS.80})
and at finite temperatures (see, for instance, Ref.
\cite{RR.94}). Since mean-field theories are variational
theories, one can carry out the projection before or after
the variation. It has turned out that variation after
projection (VAP) \cite{Zeh.67} is the appropriate tool
that fulfills the variational principle and provides
a self-consistent description of fluctuations going beyond
mean-field.

Although, the method of variation after projection has been
known since more than thirty-years, the numerical solution
of the corresponding variational equations is relatively
complicated. Therefore, a fully self-consistent variation
after exact projection has been carried out so far only in
a limited number of cases. Most of such calculations have been
restricted to the variation with respect to a few order
parameters. Applications with variation after exact
projection have been restricted to light nuclei
\cite{Schmid} or to the case of violation of particle
number in BCS-theory (see, for instance, FBCS in Ref.
\cite{DMP.64}). Exact projection within full HFB-theory 
is possible by a search of the minimum in the
projected-energy surface by gradient methods \cite{ER.82a}.
However these methods are numerically very
complicated. So far they have been applied only to the
case of number-projection in restricted spaces and for
separable-forces \cite{ER.82b}. In most of the
applications, where number-projection is necessary, the
approximate method of Lipkin and Nogami is used, which is,
in fact, only an approximation and violates the variational
principle \cite{fo96}. 

The solution of the unprojected variational problem can be
represented as a non-linear diagonalization problem (HF- or
HFB-equations). This allows an iterative solution by the
application of very fast numerical techniques. So far the
projected variational problem has not been written as a 
diagonalization problem. In this work we derive
for the first time a set of symmetry-projected HFB
equations. They have same form as in the unprojected case and
the only thing which differs is that in the case of
projection the expressions for the Hartree-Fock potential
and the pairing-field are more involved. This allows to
modify the existing codes of unprojected solutions of
the HFB-equations without much difficulity.

The manuscript is organised as follows: In section 2, some basic
HFB relations are presented. The projected HFB energy is expressed
in terms of the norm- and the Hamiltonian-overlaps in section 3
and it is shown in section 4 that these overlaps are entirely
expressible in terms of the density-matrix $\rho$ and the 
pairing-tensor $\kappa$. In section 5, it is shown that the
variation of an arbitrary real energy functional, which can
be completely written in terms of the density-matrix and
the pairing-tensor, results in HFB equations of the
conventional form. The explicit expressions of the
projected HFB fields are derived for the case of
particle-number projection in section 6. In section 7, it
is shown that the expression for the pairing-field reduces
to the familiar form of FBCS in the canonical
representation. A model study is presented in section 8 and
finally the present work is summarised in section 9.

\section{Basic HFB-Formulae}

In second quantization, the many-body Hamiltonian of a
fermion system is usually expressed in terms of a set of
annihilation and creation operators
$(c_{{}}^{{}},c_{{}}^{\dagger })=(c_{1}^{{}},\ldots
,c_{M}^{{}};c_{1}^{\dagger },\ldots ,c_{M}^{\dagger })$:
\begin{equation}
H=\sum_{n_{1}n_{2}}e_{n_{1}n_{2}}^{{}}c_{n_{1}}^{\dagger }c_{n_{2}}^{{}}+%
\frac{1}{4}\sum_{n_{1}n_{2}n_{3}n_{4}}\overline{v}%
_{n_{1}n_{2}n_{3}n_{4}}^{{}}c_{n_{1}}^{\dagger }c_{n_{2}}^{\dagger
}c_{n_{4}}^{{}}c_{n_{3}}^{{}},  
\label{E1}
\end{equation}
where the anti-symmetrized two-body matrix-element is defined by 
\begin{equation}
\overline{v}_{n_{1}n_{2}n_{3}n_{4}}^{{}}=\langle
n_{1}n_{2}|V|n_{3}n_{4}-n_{4}n_{3}\rangle .  
\label{E2}
\end{equation}
The operators annihilate the bare fermion vacuum 
\begin{equation}
c_{n}|-\rangle =0  
\label{E3}
\end{equation}
for all $n$ and provide a basis in the present analysis.

In the next step, a set of quasiparticle operators 
$(\alpha,\alpha_{}^{\dagger })=
(\alpha _{1}^{},\ldots ,\alpha_{M}^{};
\alpha _{1}^{\dagger },\ldots ,\alpha
_{M}^{\dagger })$ is introduced, which are connected to the
original particle operators by a linear transformation

\begin{eqnarray}
\quad\quad \alpha_{k}^{}
&=&\sum_{n}\left( U_{nk}^{\ast }c_{n}^{{}}+V_{nk}^{\ast}c_{n}^{\dagger }\right)
\label{E4} \\
\quad\quad \alpha_{k}^{\dagger}
&=&\sum_{n}\left( V_{nk}^{{}}c_{n}^{{}}+U_{nk}^{{}}c_{n}^{\dagger }\right) .  
\label{E5}
\end{eqnarray}
This can be written in matrix form as

\begin{equation}
\left( 
\begin{array}{c}
\alpha ^{{}} \\ 
\alpha ^{\dagger }
\end{array}
\right) =\left( 
\begin{array}{cc}
U^{\dagger } & V^{\dagger } \\ 
V^{T} & U^{T}
\end{array}
\right) \left( 
\begin{array}{c}
c \\ 
c^{\dagger }
\end{array}
\right) ={\cal W}^{\dagger }\left( 
\begin{array}{c}
c \\ 
c^{\dagger }
\end{array}
\right) .  
\label{E6}
\end{equation}
Quasiparticle operators have to satisfy the same fermion
commutation relations as the original operators. Therefore,
the transformation matrix is required to be unitary
\begin{equation}
{\cal W}^{\dagger }{\cal W}={\cal W}{\cal W}^{\dagger }=I,  
\label{E7}
\end{equation}
which leads to following relations among the coefficients 
$U_{nk}$ and $V_{nk}$ 
\begin{eqnarray}
\quad\quad U^{\dagger }U+V^{\dagger }V=I,\quad\quad 
&&UU^{\dagger }+V^{\ast }V^{T}=I,
\label{E8} \\
\quad\quad U^{T}V+V^{T}U=0,\quad\quad 
&&UV^{\dagger }+V^{\ast }U^{T}=0.  
\label{E9}
\end{eqnarray}
The quasiparticle operators annihilate the quasiparticle
vacuum $|\Phi
\rangle ,$ defined by 
\begin{equation}
\alpha _{k}|\Phi \rangle =0,  
\label{E10}
\end{equation}
for all $k$. In mean-field theory, it represents an
approximation to the ground-state of the system and turns
out to be a generalized Slater-determinant \cite{RS.80}.
Since the quasiparticle transformation mixes creation and
annihilation operators, $|\Phi \rangle $ does not
correspond to a wavefunction with good particle-number. We
therefore have two types of densities, the normal density
$\rho $, and the pairing-tensor $\kappa $, defined as
\begin{equation}
\rho _{nn^{\prime }}~=~
\langle \Phi |c_{n^{\prime }}^{\dagger }c_{n}^{{}}|\Phi \rangle ,
\quad\quad\kappa _{nn^{\prime }}~=~
\langle \Phi |c_{n^{\prime }}^{{}}c_{n}^{{}}|\Phi \rangle .  
\label{E11}
\end{equation}
These can be expressed in terms of the HFB coefficients as 
\begin{equation}
\rho =V^{\ast }V^{T},\quad\quad
\kappa =V^{\ast }U^{T}=-UV^{\dagger }.
\label{E12}
\end{equation}
It can be immediately seen that $\rho $ is Hermitian and
$\kappa $ is antisymmetric, i.e.
\begin{equation}
\rho ^{\dagger }=\rho ,\quad\quad\kappa ^{T}=-\kappa .  
\label{E13}
\end{equation}
Using the \ unitarity relations (\ref{E7}), it can be also
shown that $\rho $ and $\kappa $ satisfy
\begin{equation}
\rho -\rho ^{2}=-\kappa \kappa ^{\ast },\quad\quad
\rho \kappa =\kappa \rho^{\ast }.  
\label{E14}
\end{equation}
With these basic relations for $\rho $ and $\kappa $, the
generalized density-matrix in the Valatin form \cite{val61}
is defined as
\begin{equation}
{\cal R}=\left( 
\begin{array}{cc}
\rho & \kappa \\ 
-\kappa ^{\ast } & -\sigma ^{\ast }
\end{array}
\right) =\left( 
\begin{array}{cc}
\rho & \kappa \\ 
-\kappa ^{\ast } & 1-\rho ^{\ast }
\end{array}
\right) ={\cal W}\left( 
\begin{array}{cc}
0 & 0 \\ 
0 & 1
\end{array}
\right) {\cal W}^{\dagger },  
\label{E15}
\end{equation}
which satisfies, 
\begin{equation}
{\cal R}^{2}={\cal R}.  
\label{E16}
\end{equation}
It turns out that there is a one-to-one correspondence
between the HFB-wavefunction $|\Phi \rangle $ and the
generalized density-matrix ${\cal R}$, and that the
quasiparticle transformation ${\cal W}$ diagonalizes this
matrix. Using Thouless theorem and restricting the
following considerations to the case of even number-parity,
we can express $|\Phi \rangle $ in terms of the operators
$(c_{{}}^{{}},c_{{}}^{\dagger })$ as
\begin{equation}
|\Phi \rangle ={\cal N}e^{\hat{Z}}|-\rangle ,  
\label{E18}
\end{equation}
with 
\begin{equation}
\hat{Z}=\sum_{n<n^{\prime }}Z_{nn^{\prime }}c_{n}^{\dagger }c_{n^{\prime
},}^{\dagger }
\label{E19}
\end{equation}
where the matrix $Z$ is given by
\begin{equation}
Z=V^{\ast }U^{\ast -1}=-\rho \kappa ^{\ast -1}.
\label{E20}
\end{equation}
${\cal N}$ is a normalization factor 
\begin{equation}
{\cal N}=\sqrt{|\det U|}=\det (1-Z^{\ast }Z)^{-\frac{1}{4}}.
\label{E21}
\end{equation}

\section{The Projected-Energy}

In this section, we express the projected-energy in terms
of the Hamiltonian- and the norm-overlaps. Restricting
ourselves in this work to abelian symmetry groups, the
projection-operator is defined as
\begin{equation}
P^{I}=\int dg\,d^{I}(g)\hat{R}(g),  
\label{E22}
\end{equation}
where the integral runs over all elements $g$ of the symmetry group.
In the case of number-projection $g=\phi $ is the
gauge-angle, $I=N$ is the particle-number and 
\begin{equation}
\hat{R}(\phi )=e^{i\phi \hat{N}},\quad\quad  
d^{N}(\phi )=\frac{1}{2\pi }e^{-i\phi N}.
\label{E23}
\end{equation}
For one-dimensonal angular-momentum projection of an
axially-symmetric nucleus, $g=\beta $ is the Euler-angle,
$I$ is the angular momentum and
\begin{equation}
\hat{R}(\beta )=e^{i\beta \hat{J}_{y}},\quad\quad  
d^{I}(\beta )=\frac{2I+1}{8\pi^{2}}d_{00}^{I}(\beta ).
\label{E24}
\end{equation}
In the case of $K$-projection in triaxial nuclei, i.e. 
projection onto a good angular mometum component along 
the $z$-axis $g=\gamma $ is the azimuth angle around 
the $z$-axis, $I=K$ and 
\begin{equation}
\hat{R}(\gamma )=e^{i\gamma \hat{J}_{z}},\quad\quad
d^{K}(\gamma )=\frac{1}{2\pi}e^{-i\gamma K}.
\label{E25}
\end{equation}
Assuming that the Hamiltonian $H$ commutes with the
symmetry operator $\hat R(g)$, the projected-energy is
given by
\begin{equation}
E^{I}=\frac{\left\langle \Phi |HP^{I}|\Phi \right\rangle }
{\left\langle \Phi |P^{I}|\Phi \right\rangle }=
\frac{\int dg\,d^{I}(g)\left\langle \Phi |H\hat{R}(g)|\Phi \right\rangle }
{\int dg\,d^{I}(g)\left\langle \Phi |\hat{R}(g)|\Phi \right\rangle }.  
\label{E26}
\end{equation}
In order to simplify the notation, we introduce the rotated
Slater-determinants,
\begin{equation}
|g\rangle =
\frac{\hat{R}(g)|\Phi \rangle }
{\left\langle\Phi |\hat{R}(g)|\Phi \right\rangle },
\quad\quad{\rm with}\quad |0\rangle =|\Phi \rangle\quad\quad
{\rm and}\quad \left\langle 0|g\right\rangle =1,  
\label{E27}
\end{equation}
and the coefficients 
\begin{equation}
x(g) =d^{I}(g)\left\langle \Phi |\hat{R}(g)|\Phi \right\rangle ,
\label{E28} 
\end{equation}
and 
\begin{equation}
y(g) =\frac{x(g)}{\int dg\,x(g)},
\quad\quad{\rm with}\quad\int dg\,y(g)=1.  
\label{E29}
\end{equation}
These have the property 
\begin{equation}
x(-g)=x^{\ast }(g),\;\;\;\;y(-g)=y^{\ast }(g).  
\label{E30}
\end{equation}
Here $(-g)$ labels the inverse rotation $\hat{R}(-g)=$
$\hat{R}^{\dagger }(g).$
For the projected energy we then find 
\begin{equation}
E^{I}=\frac{\left\langle \Phi |HP^{I}|\Phi \right\rangle }{\left\langle \Phi
|P^{I}|\Phi \right\rangle }=\frac{\int dg\,x(g)\left\langle
0|H|g\right\rangle }{\int dg\,x(g)}=\int dg\,y(g)\left\langle
0|H|g\right\rangle .  
\label{E31}
\end{equation}
The overlap functions $\left\langle \Phi |\hat{R}(g)|\Phi
\right\rangle $ and $\left\langle 0|H|g\right\rangle $ are
evaluated in the following section.

\section{The Overlap-Integrals}

It will be shown in this section that the overlaps for the
norm $\langle \Phi |\hat{R}(g)|\Phi \rangle $ and the
Hamiltonian $\langle 0|H|g\rangle $ can be completely
expressed in terms of $\rho $ and $\kappa $ defined in Eq.
(\ref{E11}). For this purpose, we use the generalized Wick
theorem as derived by Onishi (for details see Ref.
\cite{RS.80}), which allows us to express overlaps in terms
the HFB coefficients $(U,V)$ and $(U_{g},V_{g})$
corresponding to the Slater determinants $|0\rangle
=|\Phi\rangle$ and $|g\rangle\sim \hat{R}(g)|\Phi \rangle$,
respectively. The HFB-coefficients $(U_{g},V_{g})$ of the
wavefunction $|g\rangle$ are given by
\begin{equation}
U_{g}^{}=D_{g}^{}U,\quad\quad  V_{g}^{}=D_{g}^{\ast }V,
\label{E32}
\end{equation}
where the matrix $D_{g}$ is the representation of the
rotation operator $\hat{R}(g)$ in the single-particle space
characterized by the basis states $|n\rangle
=a_{n}^{\dagger }|-\rangle $
\begin{equation}
\left(D_{g}\right)_{nn^{\prime }}=
\langle n|\hat{R}(g)|n^{\prime }\rangle .
\label{E33}
\end{equation}
This means 
\begin{equation}
{\cal W}_{g}={\cal D}_{g}{\cal W},\quad{\rm with}\quad
{\cal D}_{g}=\;\left( 
\begin{array}{cc}
D_{g} & 0 \\ 
0 & D_{g}^{\ast }
\end{array}
\right) .  
\label{E34}
\end{equation}

In order to derive the norm-overlap, we start from the
representation (\ref {E18}) for the HFB-wave-functions and
find
\begin{equation}
\langle \Phi |\hat{R}(g)|\Phi \rangle =|\det U|\,\langle -|e^{\hat{Z}%
^{\dagger }}e^{\hat{Z}_{g}}|-\rangle\quad{\rm with}\quad
Z_{g}=V_{g}^{\ast}U_{g}^{\ast -1}.
\label{E35}
\end{equation}
We introduce the function 
\begin{equation}
G_{g}(\lambda )=\langle -|e^{\hat{Z}^{\dagger }}e^{\lambda \hat{Z}%
_{g}}|-\rangle ,
\label{E36}
\end{equation}
which obeys the differential equation 
\begin{equation}
\frac{d}{d\lambda }G_{g}(\lambda )=
\frac{1}{2} G_{g}(\lambda )\,{\rm Tr}\ln(1-\lambda
Z_{{}}^{\ast }Z_{g}^{{}}).  
\label{E37}
\end{equation}
Integrating from $\lambda =0$ to $\lambda =1$ we find 
\begin{equation}
G_{g}(\lambda )=e^{\frac{1}{2}{\rm Tr}\ln (1-\lambda Z_{{}}^{\ast
}Z_{g}^{{}})}=\pm \sqrt{\det (1-\lambda Z_{{}}^{\ast }Z_{g}^{{}})}.
\label{E38}
\end{equation}
The phase of the overlap integral remains open in this
derivation, because the logarithme is a multivalued
function determined only up to multiples of $2\pi i.$
Therefore, we are left with a phase problem. In principle,
this phase is well defined. Several methods have been
proposed in the literature \cite{HHR.83,NW.83} to determine
the phase in an arbitrary case. However, they appear
complicated and are connected with considerable numerical
effort.

Using the expression for $G_{g}(1)$ in (\ref{E35}), we obtain 
the norm-overlap 
\begin{equation}
\langle \Phi |\hat{R}(g)|\Phi \rangle  =
|\det U|\,G_{\lambda }(1)=
\pm\sqrt{\frac{\det(1-Z_{}^{\ast }Z_{g}^{})}{\det(1-Z_{}^{\ast}Z)}}= 
\pm\sqrt{\det D_{g}}\sqrt{\det(N_{g}^{})}
\label{E41}
\end{equation}
with the matrix $N_{g}$ defined as 
\begin{equation}
N_{g}=
U_{}^{\intercal}U_{g}^{\ast}+V_{}^{\intercal }V_{g}^{\ast}=
U_{{}}^{\intercal}D_{g}^{\ast }U_{}^{\ast }+
V_{{}}^{\intercal}D_{g}^{}V_{}^{\ast},  
\label{E42}
\end{equation}

For the Hamiltonian-overlap functions, $\langle
0|H|g\rangle $, we again use the generalized Wick theorem
of Onishi, which allows us to express matrix elements of
arbitrary operators between the brackets $\langle 0|$ and
$|g\rangle $ for Slater-determinants $|g\rangle $
(normalized by $\langle 0|g\rangle =1$) in terms of the
generalized densities.  The Hamitonian overlap has three
parts
\begin{eqnarray}
\left\langle 0|H|g\right\rangle &=&H(g)=H_{sp}(g)+H_{ph}(g)+H_{pp}(g), 
\\
&=&\sum_{n_1n_2}e_{n_1n_2}^{}\left\langle 0|c_{n_{1}}^{\dagger}
c_{n_{2}}^{{}}|g\right\rangle  
\\
&&+\frac{1}{2}\sum_{n_{1}n_{2}n_{3}n_{4}}
\overline{v}_{n_{1}n_{2}n_{3}n_{4}}^{}
\left\langle 0|c_{n_{1}}^{\dagger}c_{n_{3}}^{{}}|g\right\rangle 
\left\langle 0|c_{n_{2}}^{\dagger}c_{n_{4}}^{{}}|g\right\rangle  
\\
&&+\frac{1}{4}\sum_{n_{1}n_{2}n_{3}n_{4}}
\overline{v}_{n_{1}n_{2}n_{3}n_{4}}^{}
\left\langle 0|c_{n_{1}}^{\dagger}c_{n_{2}}^{\dagger}|g\right\rangle 
\left\langle 0|c_{n_{4}}^{{}}c_{n_{3}}^{}|g\right\rangle ,
\end{eqnarray}
with 
\begin{eqnarray}
H_{sp}(g)&=&
\sum_{n_1n_2}e_{n_1n_2}^{}\rho_{n_{2}n_{1}}^{}(g)=
{\rm Tr}\left( e\rho (g)\right),  
\\
H_{ph}(g)&=&
\frac{1}{2}\sum_{n_{1}n_{2}n_{3}n_{4}}
\overline{v}_{n_{1}n_{2}n_{3}n_{4}}^{}\rho_{n_3n_1}^{}(g)
\rho_{n_4n_2}^{}(g)=
\frac{1}{2}{\rm Tr}\left(\Gamma(g)\rho(g)\right) , 
\\
H_{pp}(g) &=&
\frac{1}{4}\sum_{n_{1}n_{2}n_{3}n_{4}}
\overline{v}_{n_{1}n_{2}n_{3}n_{4}}^{}
\overline{\kappa }_{n_{1}n_{2}}^{\ast }(g)
\kappa_{n_{3}n_{4}}^{}(g), 
\\
&=&-\frac{1}{2}{\rm Tr}\left(\Delta(g)\overline{\kappa}_{}^{\ast}(g)\right)=
-\frac{1}{2}{\rm Tr}\left(\overline{\Delta }_{}^{\ast}(g)\kappa (g)\right) ,
\end{eqnarray}
where we have introduced the fields

\begin{eqnarray}
\Gamma_{n_{1}n_{3}}^{}(g) &=&
\sum_{n_{2}n_{4}}\overline{v}_{n_{1}n_{2}n_{3}n_{4}}^{}
\rho _{n_{4}n_{2}}^{}(g), 
\\
\Delta_{n_{1}n_{2}}^{}(g) &=&
\frac{1}{2}\sum_{n_{3}n_{4}}\overline{v}_{n_{1}n_{2}n_{3}n_{4}}^{}
\kappa _{n_{3}n_{4}}^{}(g), 
\\
\overline{\Delta }_{n_{1}n_{2}}^{\ast }(g) &=&
\frac{1}{2}\sum_{n_{3}n_{4}}\overline{\kappa }_{n_{3}n_{4}}^{\ast }(g)
\overline{v}_{n_{3}n_{4}n_{1}n_{2}}^{},
\\
\overline{\Delta}_{n_{1}n_{2}}^{}(g) &=&
\frac{1}{2}\sum_{n_{3}n_{4}}
\overline{v}_{n_{1}n_{2}n_{3}n_{4}}^{}
\overline{\kappa }_{n_{3}n_{4}}^{{}}(g).
\end{eqnarray}
The generalized densities are given by
\begin{eqnarray}
\rho _{nn^{\prime }}^{{}}(g) &=&\langle 0|c_{n^{\prime }}^{\dagger
}c_{n}^{{}}|g\rangle =\left( V_{g}^{\ast }N_{g}^{-1}V^{\intercal }\right)
_{nn^{\prime }},  
\label{E47} \\
\kappa _{nn^{\prime }}^{{}}(g) &=&\langle 0|c_{n^{\prime
}}^{{}}c_{n}^{{}}|g\rangle =\left( V_{g}^{\ast }N_{g}^{-1}U^{\intercal
}\right) _{nn^{\prime }},  
\label{E48} \\
-\overline{\kappa }_{nn^{\prime }}^{\ast }(g) &=&\langle 0|c_{n^{\prime
}}^{\dagger }c_{n}^{\dagger }|g\rangle =\left( U_{g}^{\ast
}N_{g}^{-1}V^{\intercal }\right) _{nn^{\prime }},  
\label{E49} \\
-\sigma _{nn^{\prime }}^{\ast }(g) &=&\langle 0|c_{n^{\prime
}}^{{}}c_{n}^{\dagger }|g\rangle =\left( U_{g}^{\ast }N_{g}^{-1}U^{\intercal
}\right) _{nn^{\prime }}.  
\label{E50}
\end{eqnarray}
The matrices $\kappa (g)$ and $\overline{\kappa }(g)$ are
anti-symmetric and the matrix $\sigma (g)$ is related to
$\rho (g)$ through
\begin{equation}
\sigma (g)=\rho ^{\dagger }(g)-1.
\label{E53}
\end{equation}

So far all the expressions still depend on the HFB
coefficients $U$ and $V$.  Using the definitions for the
densities $\rho $ and $\kappa $, we now rewrite expressions
for the overlaps entirely in terms of these quantities as
\begin{eqnarray}
\langle \Phi |\hat{R}(g)|\Phi \rangle  &=&
\pm \sqrt{\det D_{g}}
\sqrt{\det (U^{T}D_{g}^{\ast} U_{}^{\ast} + V^{T}D_{g}^{\ast} V_{}^{\ast})}
\nonumber \\
&=&\pm \sqrt{\det D_{g}} 
\sqrt{\det \{ V^{-1} (V U^{\dagger}D_{g}^{\dagger} U_{}^{} V^{\dagger} + 
V V^{\dagger}D_{g}^{T} V_{}^{} V^{\dagger})(V^{\dagger})^{-1}\}}
\nonumber \\
&=&\pm \sqrt{\det D_{g}}(\det \rho
)^{-1/2}\sqrt{\det \left( \rho D_{g}^{{}}\rho -\kappa D_{g}^{\ast }\kappa
^{\ast }\right) }, \\
%&=&\pm \det D_{g}(\det \rho )^{-1/2}\sqrt{\det \left( \rho \rho _{g}-\kappa
%\kappa _{g}^{\ast }\right) },  
%\label{E55} \\
&=&\pm \frac{\det D_{g}}{\sqrt{\det \rho \det C_{g}}},  
\label{E56}
\end{eqnarray}
where we have introduced the rotated-densities 
\begin{eqnarray}
\rho _{g} &=&D_{g}^{{}}\rho D_{g}^{\dagger },
\;\;\;\rho _{-g}=D_{g}^{\dagger}\rho D_{g}^{}, 
\label{E58} 
\\
\kappa _{g} &=&D_{g}^{{}}\kappa D_{g}^{\intercal },\;\;\;\kappa
_{-g}=D_{g}^{\dagger }\kappa D_{g}^{\ast },  
\label{E59} 
\\
Z_{g} &=&D_{g}^{{}}ZD_{g}^{\intercal },\;\;\;Z_{-g}=D_{g}^{\dagger
}ZD_{g}^{\ast }
\label{E60} 
\end{eqnarray}
and 
\begin{equation}
C_{g}^{-1}=\rho \rho _{g}-\kappa \kappa _{g}^{\ast }.  
\label{E61}
\end{equation}
The transition densities appearing in the
Hamiltonian-overlaps can also be rewritten in terms of the
HFB densities as
\begin{eqnarray}
\rho (g) &=&\rho _{g}C_{g}^{{}}\rho,  
\label{E65} \\
\kappa (g) &=&\rho _{g}C_{g}^{{}}\kappa,  
\label{E66} \\
\overline{\kappa }_{{}}^{\ast }(g)&=&\kappa^{\ast}_{g} C_{g} \rho,  
\label{E67} \\
\sigma _{{}}^{\ast }(g) &=&\kappa _{g}^{\ast }C_{g}^{{}}\kappa .  
\label{E68}
\end{eqnarray}
It is now clear from the above expressions that the norm-
and the Hamiltonian- overlaps are completely expressible in
terms of densities $\rho $ and $\kappa $. In the next
section, it will be shown that the variation of an
arbitrary real energy functional which is expressible in
terms of these densities results in the HFB equations.
\section{Variational Equations}

We have now expressed the projected-energy as a functional
of the HFB-densities $\rho $ and $\kappa $. So far we have
used the fact that $\rho $ in hermitian and that $\kappa $
is skew-symmetric, but we have not used the relations
${\cal R}^{2}={\cal R}$ ($\rho -\rho ^{2}=-\kappa \kappa ^{\ast }$ and
$\rho\kappa =\kappa \rho ^{\ast })$. Without these
equations we have the independent variables $\Re e(\rho
_{n^{\prime }n}),\Im m(\rho _{n^{\prime }n}),\rho _{nn},$
$\Re e(\kappa _{n^{\prime }n}),$ and $\Im m(\kappa
_{n^{\prime }n})$ for $n<n^{\prime }$ and we have to vary
with respect to them under the constraint ${\cal
R}^{2}={\cal R}$. As shown in standard text books, this
variation can be replaced by a variation with respect to
the independent variables $\rho _{n^{\prime }n}^{{}},\rho
_{n^{\prime }n}^{\ast },\rho _{nn},\kappa _{n^{\prime
}n}^{{}},$ and $\kappa _{n^{\prime }n}^{\ast }$ for
$n<n^{\prime }$under the constraint, ${\cal R}^{2}={\cal
R}.$ Introducing the Lagrangian multipliers $\Lambda
_{nn^{\prime }}$ the variational ansatz is
\begin{equation}
\delta \left\{ E(\rho ,\kappa )-{\rm Tr}\left( \Lambda ({\cal R}^{2}-{\cal R}%
)\right) \right\} =0.
\label{E80}
\end{equation}
As we have shown in the earlier section that the projected-energy
depends only on $\rho $ and $\kappa $, the variation of the
energy is expressed as 
\begin{eqnarray}
\delta E &=&
\sum_{n<n^{\prime }}
\left( \frac{\partial E}{\partial\rho_{n^{\prime }n}^{}} 
\delta \rho_{n^{\prime }n}^{}+
\frac{\partial E}{\partial \rho _{n^{\prime }n}^{\ast }}
\delta \rho _{n^{\prime }n}^{\ast }+
\frac{\partial E}{\partial \kappa _{n^{\prime }n}^{}}
\delta \kappa_{n^{\prime }n}^{}+
\frac{\partial E}{\partial \kappa _{n^{\prime }n}^{\ast}}
\delta \kappa _{n^{\prime }n}^{\ast }\right) \nonumber \\
&+& \sum_{n}\frac{\partial E}{\partial \rho _{nn}^{}}
\delta \rho _{nn}^{},
\label{E81}
\end{eqnarray}
and introducing the quantities 
\begin{equation}
h_{nn^{\prime }}^{}=
\frac{\partial E}{\partial \rho _{n^{\prime }n}^{}}
\quad {\rm for}\quad n\leq n^{\prime },\quad
\Delta _{nn^{\prime}}^{}=
-\frac{\partial E}{\partial \kappa _{n^{\prime }n}^{\ast }}
\quad{\rm for} \quad n<n^{\prime }.
\label{E82}
\end{equation}
Since the functional $E$ is real, we find 
\begin{equation}
h_{nn^{\prime }}^{\ast }=
\frac{\partial E}{\partial \rho _{n^{\prime}n}^{\ast}},\quad\quad 
\Delta _{nn^{\prime }}^{\ast }=
-\frac{\partial E}{\partial \kappa _{n^{\prime }n}^{}}.
\label{E83}
\end{equation}
It can be easily shown that 
\begin{equation}
h_{nn^{\prime }}^{}=h_{n^{\prime }n}^{\ast },\quad
\Delta_{nn^{\prime }}^{}=-\Delta _{n^{\prime }n}^{{}},
\label{E84}
\end{equation}
and obtain 
\begin{eqnarray}
\delta E &=&\frac{1}{2}\left\{ \sum_{nn^{\prime }}\frac{\partial E}{\partial
\rho _{n^{\prime }n}^{{}}}\delta \rho _{n^{\prime }n}^{{}}+\frac{\partial E}{%
\partial \rho _{n^{\prime }n}^{\ast }}\delta \rho _{n^{\prime }n}^{\ast }+%
\frac{\partial E}{\partial \kappa _{n^{\prime }n}^{{}}}\delta \kappa
_{n^{\prime }n}^{{}}+\frac{\partial E}{\partial \kappa _{n^{\prime }n}^{\ast
}}\delta \kappa _{n^{\prime }n}^{\ast }\right\} , 
\label{E85}\\
&=&\frac{1}{2}\left\{ \sum_{nn^{\prime }}h_{nn^{\prime }}^{{}}\delta \rho
_{n^{\prime }n}^{{}}+h_{nn^{\prime }}^{\ast }\delta \rho _{n^{\prime
}n}^{\ast }-\Delta _{nn^{\prime }}^{\ast }\delta \kappa _{n^{\prime
}n}^{{}}-\Delta _{nn^{\prime }}^{{}}\delta \kappa _{n^{\prime }n}^{\ast
},\right\} , 
\label{E86}\\
&=&\frac{1}{2}\left\{ {\rm Tr}\left( h\delta \rho \right) +{\rm Tr}\left(
h^{\ast }\delta \rho ^{\ast }\right) -{\rm Tr}\left( \Delta ^{\ast }\delta
\kappa \right) -{\rm Tr}\left( \Delta \delta \kappa ^{\ast }\right) \right\}
,\label{E87} \\
&=&\frac{1}{2}{\rm Tr}\left\{ \left( 
\begin{array}{cc}
h & \Delta \\ 
-\Delta ^{\ast } & -h^{\ast }
\end{array}
\right) \left( 
\begin{array}{cc}
\delta \rho & \delta \kappa \\ 
-\delta \kappa ^{\ast } & -\delta \rho ^{\ast }
\end{array}
\right) \right\} ,
\label{E88}
\end{eqnarray}
and introducing the matrix 
\begin{equation}
{\cal H}=\left( 
\begin{array}{cc}
h & \Delta \\ 
-\Delta ^{\ast } & -h^{\ast }
\end{array}
\right) ,
\end{equation}
we have 
\begin{equation}
\delta E=\frac{1}{2}{\rm Tr}\left( {\cal H}\delta {\cal R}\right) .
\end{equation}
Including the constraint leads to the variational ansatz 
\begin{equation}
\delta \left\{ E(\rho ,\kappa )-{\rm Tr}\left( \Lambda ({\cal R}^{2}-{\cal R}%
)\right) \right\} =\frac{1}{2}{\rm Tr}\left\{ ({\cal H-}\Lambda {\cal R-R}%
\Lambda +\Lambda )\delta {\cal R}\right\} =0.
\end{equation}
Since $\delta {\cal R}$ is an arbitrary variation, we find 
\begin{equation}
{\cal H-}\Lambda {\cal R-R}\Lambda +\Lambda =0.
\end{equation}
Using $({\cal R}^{2}-{\cal R})$, the above equation can be written as 
\begin{equation}
\left[ {\cal H},{\cal R}\right] =0,
\end{equation}
which is solved by the HFB-equations 
\begin{equation}
\left( 
\begin{array}{cc}
h & \Delta \\ 
-\Delta ^{\ast } & -h^{\ast }
\end{array}
\right) \left( 
\begin{array}{c}
U \\ 
V
\end{array}
\right) =\left( 
\begin{array}{c}
U \\ 
V
\end{array}
\right) E.  \label{E.75}
\end{equation}
The HFB-equations have been obtained in (\ref{E.75}) with the only
assumption that the energy functional is expressible in terms of $\rho $ and 
$\kappa$. As we have shown that the projected-energy $E^{I}$ 
is also completely expressible
in terms of these quantities. Therefore, in the case of variation 
after projection we find
HFB-equations of the same structure. We only have to consider a 
projected-energy functional 
\begin{equation}
E^{I}=\frac{\left\langle \Phi |HP^{I}|\Phi \right\rangle }{\left\langle \Phi
|P^{I}|\Phi \right\rangle }=\int dg\,y(g)\left\langle 0|H|g\right\rangle ,
\end{equation}
which yields different expressions for the fields $h$ and $\Delta $: 
\begin{equation}
h_{nn^{\prime }}^{I}=\frac{\partial E^{I}}{\partial \rho _{n^{\prime }n}^{{}}%
},\;\;\;\;\;\;\;\;\;\;\Delta _{nn^{\prime }}^{I}=-\frac{\partial E^{I}}{%
\partial \kappa _{n^{\prime }n}^{\ast }}.
\end{equation}
In the following section, these derivatives are obtained for the case of
particle-number projection. The expressions of these derivatives in a
general case become quite complicated and will not be presented here.

\section{Projection of Particle-Number}

Number-projection is a simple example of the projection theory since 
in this case the matrix $%
D_{g} $ is just a phase factor, a multiple of a unit matrix. The projection
operator which projects out the
good particle number $(N)$ is of the form 
\begin{equation}
P^{N}=\frac{1}{2\pi }\int d\phi \,e^{i\phi (\hat{N}-N)}.
\end{equation}
The rotation in the gauge-space is given by
\begin{equation}
D_{\phi }=e^{i\phi }.
\end{equation}
Using this definition in Eqs. (\ref{E58}) and (\ref{E59}), we obtain the
expressions for the rotated-densities in the case of particle-number
projection as
\begin{equation}
\rho _{\phi }^{{}}=\rho ,\;\;\;\;\kappa _{\phi }^{{}}=e^{2i\phi }\kappa.
\end{equation}
This simplifies the expressions for all the matrices defined in section IV.
In particular, we have 
\begin{equation}
C_{\phi}^{-1}\equiv
\rho\rho^{}_\phi-\kappa\kappa_\phi^\ast
=e^{-i\phi }\rho
\left( e^{i\phi }\rho +e^{-i\phi }(1-\rho )\right) =\rho C^{-1}(\phi )
\end{equation}
with 
\begin{eqnarray}
C(\phi ) &=&\rho^{-1} C_{\phi }=\left( \rho +e^{-2i\phi }(1-\rho )\right)
^{-1} \\
&=&e^{2i\phi }\left( 1+\rho (e^{2i\phi }-1)\right) ^{-1}
\label{E201}
\end{eqnarray}
and 
\begin{equation}
C_{{}}^{\dagger }(\phi )=C(-\phi ).
\end{equation}
For the transition densities we obtain 
\begin{eqnarray}
\rho (\phi ) &=&C(\phi )\rho, \label{202}\\
\kappa (\phi ) &=&C(\phi )\kappa =\kappa C_{{}}^{\intercal }(\phi ),
\label{E203} \\
\overline{\kappa }(\phi ) &=&e^{2i\phi }\kappa C_{{}}^{\ast }(\phi
)=e^{2i\phi }C_{{}}^{\dagger }(\phi )\kappa,\label{204}
\end{eqnarray}
and 
\begin{eqnarray}
\rho (-\phi ) &=&C_{{}}^{\dagger }(\phi )\rho =\rho _{{}}^{\dagger }(\phi ),
\\
\kappa (-\phi ) &=&e^{-2i\phi }\overline{\kappa }(\phi )=C_{{}}^{\dagger
}(\phi )\kappa =\kappa C_{{}}^{\ast }(\phi ), \\
\overline{\kappa }(-\phi ) &=&e^{-2i\phi }\kappa (\phi )=e^{-2i\phi }\kappa
C_{{}}^{\intercal }(\phi )=e^{-2i\phi }C(\phi )\kappa
\end{eqnarray}
Considering that $\det D_{\phi }=e^{iM\phi }$, we find for the norm-overlap 
\begin{equation}
x(\phi )=\frac{1}{2\pi }e^{-i\phi N}\sqrt{\det \left( e^{2i\phi }\rho
+(1-\rho )\right) }=\frac{1}{2\pi }\frac{e^{i\phi (M-N)}}{\sqrt{\det C(\phi )%
}}.
\end{equation}
For the Hamiltonian-overlap, we have 
\begin{eqnarray}
H_{sp}(\phi ) &=&{\rm Tr}\left( e \rho(\phi) \right), \\
H_{ph}(\phi ) &=&\frac{1}{2}{\rm Tr}\left( \Gamma (\phi )\rho(\phi)\right) =%
\frac{1}{2}\sum_{n_{1}n_{2}n_{3}n_{4}}(\rho(\phi))_{n_{3}n_{1}}^{{}}
\overline{v}_{n_{1}n_{2}n_{3}n_{4}}^{{}}(\rho(\phi))_{n_{4}n_{2}}^{{}}, \\
H_{pp}(\phi ) &=&-\frac{1}{2}{\rm Tr}\left(\overline{\Delta }_{{}}^{\ast}
(\phi)\kappa(\phi)\right) =-\frac{1}{2}{\rm Tr}\left( \Delta (\phi)
\overline{\kappa}_{{}}^{\ast}(\phi)\right), \\
&=&\frac{1}{4}\sum_{n_{1}n_{2}n_{3}n_{4}}
\overline{\kappa}^{\ast}_{n_{1}n_{2}}(\phi)
\overline{v}_{n_{1}n_{2}n_{3}n_{4}}^{{}}\kappa_{n_{3}n_{4}}^{}(\phi),
\end{eqnarray}
with

\begin{eqnarray}
\Gamma _{n_{1}n_{3}}^{{}}(\phi ) 
&=&\sum_{n_2n_4}\overline{v}_{n_1n_2n_3n_4}^{}
\rho_{n_4n_2}^{}(\phi),\\
\Delta _{n_1n_2}^{}(\phi ) 
&=&\frac{1}{2}\sum_{n_3n_4}\overline{v}_{n_1n_2n_3n_4}^{}
\kappa_{n_3n_4}^{}(\phi), \\
\overline{\Delta}_{n_3n_4}^{\ast}(\phi ) 
&=&\frac{1}{2}\sum_{n_1n_2}\overline{\kappa}^{\ast}_{n_1n_2}(\phi)
\overline{v}_{n_{1}n_{2}n_{3}n_{4}}^{{}},
\end{eqnarray}
with 
\begin{eqnarray}
\Gamma (-\phi ) &=&\Gamma _{{}}^{\dagger }(\phi ), \\
\Delta (-\phi ) &=&e^{-2i\phi }\overline{\Delta }(\phi ), \\
\overline{\Delta }(-\phi ) &=&e^{-2i\phi }\Delta (\phi ).
\end{eqnarray}
Summarizing, we can write the projected-energy as 
\begin{equation}
E^{N}=\int d\phi \,y(\phi )\left( H_{sp}(\phi )+H_{ph}(\phi )+H_{pp}(\phi
)\right) . 
\end{equation}
In the following subsections, we shall evaluate the variation of the
norm- and the Hamiltonian-overlaps.

\subsection{Variation of the Norm}

Using (\ref{E14}), the norm in the case of the particle-number 
projection can be rewritten
in terms of the density $(\rho)$ only. 
This simplifies the variational
equations considerably. Defining the matrices $X(\phi)$ and $Y(\phi)$ by
\begin{eqnarray}
\frac{\partial x(\phi)}{\partial\rho_{n^\prime n}} 
&=&x(\phi) X_{nn^\prime}(\phi),
\\
\frac{\partial y(\phi)}{\partial\rho_{n^\prime n}} 
&=&y(\phi) Y_{nn^\prime}(\phi),
\end{eqnarray}
we find by differentiation
\begin{eqnarray}
\frac{\partial x(\phi )}{\partial \rho _{n^{\prime }n}} 
&=& \frac{1}{2}x(\phi ){\rm Tr}\left(C(\phi)
\frac{\partial}{\partial\rho_{n^{\prime }n}}C_{{}}^{-1}(\phi)\right), \\
&=&\frac{1}{2}x(\phi )(1-e^{-2i\phi })C_{nn^{\prime }}(\phi)
~=~x(\phi )ie^{-i\phi }\sin\phi\;C_{nn^{\prime }}(\phi ),
\end{eqnarray}
and obtain 
\begin{equation}
X(\phi )=ie^{-i\phi}\sin\phi\;C(\phi ),
\end{equation}
and 
\begin{equation}
Y(\phi )=\,ie^{-i\phi}\sin\phi\;C(\phi )-i
\int d\phi ^{\prime }y(\phi^\prime)e^{-i\phi^\prime}\sin\phi^\prime\;
C(\phi^\prime).
\end{equation}

\subsection{The Projected Hartree-Fock Field}

The HF-field in the projected HFB-equtions is obtained as the derivative
of the number-projected energy with respect to the density
\begin{eqnarray}
h^N_{nn^{\prime }}&=&\frac{\partial E^{N}}{\partial \rho _{n^{\prime}n}}\\
&=&\int d\phi\,y(\phi)\left(Y_{nn^\prime}(\phi)H(\phi)+ 
\frac{\partial H(\phi)}{\partial\rho _{n^\prime n}}\right)
\quad\quad\quad{\rm for}\quad (n\leq n^{\prime }),
\end{eqnarray}
which gives 
\begin{eqnarray}
h^N &=&\frac{\partial E^N}{\partial\rho}=
\int d\phi \,\,y(\phi )Y(\phi)(H_{sp}(\phi)+H_{ph}(\phi)+H_{pp}(\phi)) 
\\
&&\quad\quad\quad+\int d\phi\;y(\phi)\frac{\partial}{\partial\rho} 
\left(H_{sp}(\phi)+H_{ph}(\phi)+H_{pp}(\phi)\right). 
\end{eqnarray}
The above projected HF-potential has three parts 
\begin{equation}
h^N  = \varepsilon^N + \Gamma^N + \Lambda^N
\end{equation}
with
\begin{eqnarray}
\varepsilon^N &=& 
\int d\phi \,\,y(\phi )\left(Y(\phi)H_{sp}(\phi)+
\frac{\partial}{\partial\rho} H_{sp}(\phi)\right),
\\
\Gamma^N &=&
\int d\phi \,\,y(\phi )\left(Y(\phi)H_{ph}(\phi)+
\frac{\partial}{\partial\rho} H_{ph}(\phi)\right),
\\
\Lambda^N &=&
\int d\phi \,\,y(\phi )\left(Y(\phi)H_{pp}(\phi)+
\frac{\partial}{\partial\rho} H_{pp}(\phi)\right).
\end{eqnarray}
Using 
\begin{equation}
{\rm Tr}\left( A\frac{\partial }{\partial\rho_{n^{\prime }n}}C(\phi)\right)
=-2ie^{-i\phi}\sin\phi\;
\left(C(\phi)AC(\phi)\right)_{nn^{\prime }},
\end{equation}
we find 
\begin{eqnarray}
\frac{\partial H_{sp}(\phi )}{\partial \rho _{n^{\prime }n}} 
&=&{\rm Tr}\left( e\frac{\partial }{\partial \rho _{n^{\prime }n}}
\left[ C(\phi )\rho \right] \right) 
\quad\quad\quad\quad\quad{\rm for}\quad (n\leq n^{\prime }) 
\\
&=&\left([1-2ie^{-i\phi}\sin\phi\rho(\phi)]
eC(\phi)\right)_{nn^{\prime }},
\\
\frac{\partial H_{ph}(\phi )}{\partial\rho_{n^{\prime }n}} 
&=&{\rm Tr}\left( \Gamma (\phi)\frac{\partial}{\partial\rho _{n^{\prime}n}}
\left[C(\phi )\rho\right]\right)
\quad\quad\quad\quad{\rm for}\quad (n\leq n^{\prime }) 
\\
&=&\left([1-2ie^{-i\phi}\sin\phi\rho(\phi)]
\Gamma(\phi)C(\phi )\right)_{nn^{\prime}},
\\
\frac{\partial H_{pp}(\phi)}{\partial \rho _{n^{\prime }n}} 
&=&-\frac{1}{2}{\rm Tr}\left(\overline{\Delta }^{\ast }(\phi )
\frac{\partial }{\partial\rho _{n^{\prime }n}}\left[ C(\phi )\kappa \right] 
\right) \nonumber \\ 
&-&\frac{1}{2} e^{-2i\phi }{\rm Tr}\left( \Delta (\phi )\frac{\partial }
{\partial\rho_{n^{\prime }n}}\left[ \kappa _{{}}^{\ast }C\right] 
\right)  
\quad\quad\quad{\rm for}\quad (n\leq n^{\prime })
\\
&=&ie^{-i\phi}\sin\phi\;\left(
\kappa(\phi)\overline{\Delta }^{\ast }(\phi)C(\phi )+
C(\phi)\Delta(\phi)\overline{\kappa}_{}^{\ast}(\phi)\right)_{nn^{\prime}}.
\end{eqnarray}
Considering Hermiticity, we finally have
\begin{eqnarray}
\varepsilon^N &=& 
\frac{1}{2}\int d\phi \,\,y(\phi )\biggl( Y(\phi)
{\rm Tr}[e\rho(\phi)] \nonumber \\
&+& [1-2ie^{-i\phi}\sin\phi\rho(\phi)] e C(\phi)\biggr) ~+~h.c.
\\
\Gamma^N &=&
\frac{1}{2}\int d\phi \,\,y(\phi )\biggl( Y(\phi)
\frac{1}{2}{\rm Tr}[\Gamma(\phi)\rho(\phi)] \nonumber \\ 
&+& \frac{1}{2} [1-2ie^{-i\phi}\sin\phi\rho(\phi)]\Gamma(\phi)C(\phi)\biggr)~+~h.c.
\\
\Lambda^N &=&
-\frac{1}{2}\int d\phi \,\,y(\phi )\biggl( Y(\phi)
\frac{1}{2}{\rm Tr}[\Delta(\phi)\overline{\kappa}_{}^\ast(\phi)] \nonumber \\
&-& 2 ie^{-i\phi}\sin\phi\; C(\phi)\Delta(\phi)
\overline{\kappa}_{}^{\ast}(\phi)\biggl)
~+~h.c.
\end{eqnarray}

\subsubsection{The Projected-Pair Field}

The pairing-field is obtained by variation of the projected-energy
with respect to $\kappa$
\begin{eqnarray}
\Delta^N_{nn^{\prime }}&=&
-\frac{\partial E^{N}}{\partial\kappa^\ast_{n^\prime n}}
\quad\quad\quad{\rm for}\quad (n <  n^{\prime })\\
&=&-\int d\phi\,y(\phi)
\frac{\partial H(\phi)}{\partial\kappa^\ast_{n^\prime n}}.
\end{eqnarray}
We find
\begin{eqnarray}
\frac{\partial H_{pp}(\phi )}{\partial \kappa _{n^{\prime }n}^{\ast }} 
&=&-\frac{1}{2}e^{-2i\phi }{\rm Tr}\left(C(\phi)\Delta (\phi)
\frac{\partial }{\partial\kappa_{n^\prime n}^\ast}\kappa_{}^\ast\right) 
\quad\quad\quad{\rm for}\quad (n <  n^{\prime }) 
\\
&=&-\frac{1}{2}e^{-2i\phi }\left( C(\phi )\Delta (\phi )-(..)^{\intercal
}\right) _{nn^{\prime }},
\end{eqnarray}
which finally yields the pairing-field in the 
projected HFB-equations 
\begin{equation}
\Delta^N =\frac{1}{2}\int d\phi e^{-2i\phi }\;y(\phi )C\left( \phi \right) \Delta
(\phi )-(..)^{\intercal }
\label{E301}  
\end{equation}

\section{Canonical representation}

In this section, we shall demonstrate that the projected expression for the $%
\Delta $-matrix given in the last section reduces to the familiar form in the
canonical basis. In the canonical basis, which we shall denote by the greek
indices $(\mu ,\nu ,...)$, the density-matrix and the pairing-tensor reduce
to the following $(2\times 2)$ matrices 
\[
\rho _{\mu }=\left( 
\begin{array}{cc}
v_{\mu }^{2} & 0 \\ 
0 & v_{\mu }^{2}
\end{array}
\right) ,~~~~~~~~\kappa _{\mu }=\left( 
\begin{array}{cc}
0 & u_{\mu }v_{\mu } \\ 
-u_{\mu }v_{\mu } & 0
\end{array}
\right) . 
\]
Using these basic expressions, the matrices in the number-projection,
Eqs. (\ref{E201}-\ref{E203}) acquire the
following form in the canonical basis 
\begin{eqnarray}
C_{\mu }(\phi ) &=&\frac{1}{u_{\mu }^{2}+e^{2\imath \phi }v_{\mu }^{2}}=%
\frac{1}{u_{\mu }^{2}+\zeta v_{\mu }^{2}} \label{E401}\\
x(\phi ) &=&\frac{e^{-\imath N\phi }}{2\pi }\prod_{\mu >0}\left( u_{\mu
}^{2}+e^{2\imath \phi }v_{\mu }^{2}\right) \\
&=&\frac{1}{2\pi }\zeta ^{-n}\prod_{\mu >0}\left( u_{\mu }^{2}+\zeta v_{\mu
}^{2}\right) \label{E402}\\
\rho _{\mu }(\phi ) &=&e^{\imath \phi }C_{\mu }(\phi )v_{\mu }^{2}\times I
\label{E403} \\
\kappa _{\mu }(\phi ) &=&e^{\imath \phi }C_{\mu }(\phi )u_{\mu }v_{\mu
}\times {\cal I} \label{404}\\
\bar{\kappa}_{\mu }^{\ast }(\phi ) &=&e^{-\imath \phi }C_{\mu }(\phi )u_{\mu
}v_{\mu }.\times {\cal I} \label{E405},
\end{eqnarray}
where $\zeta =e^{2i\phi }$ and %
\begin{equation}
I=\left( 
\begin{array}{cc}
1 & 0 \\ 
0 & 1
\end{array}
\right) \;\;\;\;{\rm and\;\;\;\;}{\cal I}=\left( 
\begin{array}{cc}
0 & 1 \\ 
-1 & 0
\end{array}
\right).
\end{equation}
The norm is therefore given by 
\begin{equation}
<\Phi |P^{N}|\Phi >=\int_{0}^{2\pi }d\phi \,x(\phi )=\frac{1}{2\pi \imath }%
\oint d\zeta ~~\zeta ^{-n-1}\prod_{\mu >0}\left( u_{\mu }^{2}+\zeta v_{\mu
}^{2}\right) =R_{0}^{0},
\end{equation}
where the residuum interals are defined in analogy to ref. \cite{DMP.64}: 
\begin{equation}
R_{n}^{m}(\mu _{1},\dots ,\mu _{m})=\frac{1}{2\pi \imath }\oint d\zeta
~~\zeta ^{m-n-1}\prod_{\mu \neq \mu _{1}..\mu _{m}}\left( u_{\mu }^{2}+\zeta
v_{\mu }^{2}\right) .
\end{equation}
Using the expression for the pair-gap (\ref{E301})
and the canonical forms of the matrices in (\ref{E401}-\ref{E405}), we obtain 
\begin{eqnarray}
\Delta _{\mu \bar{\mu}^{\prime }} &=&\frac{1}{8\pi \imath }\frac{1}{<\Phi
_{0}|P^{N}|\Phi _{0}>}\times \\
\oint d\zeta ~~\zeta ^{-N}\sum_{\nu } &<&\mu \bar{\mu}^{\prime }|v|\nu \bar{%
\nu}>\frac{\prod_{\nu ^{\prime }}(\zeta ^{\ast }u_{\nu ^{\prime }}^{2}+\zeta
v_{\nu ^{\prime }}^{2})}{(\zeta ^{\ast }u_{\mu }^{2}+\zeta v_{\mu
}^{2})(\zeta ^{\ast }u_{\nu }^{2}+\zeta v_{\nu }^{2})}  \nonumber \\
&-&(\mu \leftrightarrow \mu ^{\prime })
\end{eqnarray}
The diagonal form $(\mu =\mu ^{\prime })$ of this expression agrees
with the expression derived in ref.\cite{DMP.64}. The expresions for the
other fields in the canonical representation also agree with the 
expressions given in ref.\cite{DMP.64}. 

\section{Model study}
As a case study for the projection formalism developed in the present work, 
we use the deformed shell model Hamiltonian which 
consists of a deformed
one-body term, $h$ and a scalar two-body delta-interaction 
\cite{snrp89}. The one-body term is the familiar
Nilsson mean-field potential which takes into account of the
long-range part of the nucleon-nucleon interaction. The residual short-range
interaction is specified by the delta-interaction. 
The model Hamiltonian employed is given by
\begin{equation} \label{H}
H = -4 \kappa { \sqrt { 4 \pi \over 5 }} Y_{20} -g \delta(\hat r_1 - \hat r_2).
\label{E501}
\end{equation}
We use 
$G=g\int R^4_{nl}r^2 dr$ as our energy unit and the deformation
energy $\kappa$ is related to the deformation parameter $\beta$.
For the case of $h_{11/2}$ shell, $\kappa$=2.4 approximately
corresponds to $\beta=0.25$. The model Hamiltonian has been solved 
exactly for particles in $h_{11/2}$ intruder subshell and the results
will be compared with those obtained using the projection formalism.
In the numerical calculations, we have used only $J=0$ and 2 components
of the $\delta$-interaction. The problem with using the full 
$\delta$-interaction is that in the HFB analysis of this simple model,
the contribution of the higher-multipoles to the particle-hole channel is
quite large and it renormalises the one-body potential substantially. 
This renormalization makes it very difficuilt to choose a reasonable
initial HFB fields. In most of the mean-field analysis this particle-hole
channel is not considered. However, since we are comparing the results
with the exact calculations, it is important to keep all the components
of the mean-field. Nevertheless, the main purpose here is to choose
a simple model which can solved exactly and the projection
method will be tested using these exact results.

In the HFB analysis of the single-j shell, the single-particle basis are
the magnetic-states of the single-particle angular-momentum, $j$. 
The summation indices in all the expressions given in 
section VI run over these magnetic-states.
For the case of
$h_{11/2}$ orbital, the summation indices have the range $(-11/2,....,11/2)$
and since the Hamiltonian in (\ref{E501}) obeys the time-reversal symmetry, it is 
required to have only +ive
or -ive magnetic-states. Therefore, the summation indices have dimensions
of six for the $h_{11/2}$ orbital. 

In order to test the projection 
method, we have
performed HFB, PHFB and the exact shell model calculations as a function
of the pairing strength, $G$ for 6-particles in $h_{11/2}$. In Fig. 1, 
the results of total  energy ($E_{tot}$) and pairing energy ($E_{pair}$)
are presented. In PHFB we have used six-mesh points for integration over
the gauge-angle. It is found that for the value of $G$ close to 1,
the three mesh-points give quite accurate results. However, for the 
limiting value
of $G$ close to 0, it is important to use six-mesh points. The PHFB results
of $E_{tot}$ with six-mesh points reproduce the shell model results
almost exactly. In Fig. 1 the results of PHFB and the shell model
calculations are indistinguishable. As expected
the results of PHFB and HFB are identical for $G=0$ since there is no
two-body interaction and $E_{tot}$ is equal to the energy of the one-body
static-potential. The results of PHFB and HFB deviate with increasing value
of $G$ and the deviation is more than 1 unit for the value of
$G=1$. The results of $E_{pair}$ are shown in the lower pannel of Fig. 1 and
it is clear that HFB shows a phase transition at $G=0.4$. For $G=0.4$ and
lower, $E_{pair}$ is exactly zero. The PHFB on the other hand depicts
no such phase transition and has a finite value for finite $G$. 

\section{Summary}

In order to obtain an approximate solution of the many-body problem, it
is often required to break the symmetries which the original 
many-body Hamiltonian obeys. For instance, in the case of
Hartree-Fock or 
Hartree-Fock-Bogoliubov approaches, the many-body
wavefunction is approximated by a Slater-determinant which in
a general case breaks the rotational and the particle-number symmetries. In 
order to have a better description for the many-body problem, it is essential
to restore the broken-symmetries.  

The symmetry restoration may not be very critical in strongly symmetry
breaking conditions, but for the case of a weakly broken symmetry the
restoration is quite important. For example, in the case of well deformed
nuclei in the ground state, the pair-correlations are strong
and the BCS theory which breaks the particle-number symmetry, describes the
pair-correlation quite reasonably. However, with increasing
quasiparticle excitation, the pair-correlations reduce and 
the BCS solution collapses. The pair-correlations are also quenched
for superdeformed shapes. For these weak pairing cases, it is quite
essential to consider the restoration of the particle-number. As already
mentioned in the introduction, the most undesirable feature
associated with the mean-field theory is that one obtains unphysical
phase transitions, for instance from finite-pairing to zero-pairing.
These phase transitions are smeared out with the restoration of the
broken symmetries. 

In most cases, the projection methods have been carried out approximately
\cite{fo96}.
A self-consistent description 
of the projection in the HFB framework has been an unresolved problem.
In the present work, we have obtained the
symmetry-projected HFB equations for the first time. This has been 
possible by first
realising that the projected-energy can be completely expressed
in terms of the HFB densities and then it has been demonstrated that
the variation of an arbitrary energy functional which is completely
expressible in terms of HFB densities, results in HFB equations. The
expressions for the Hartree-Fock and the pair-field depend on the
form of the energy functional. In the case of the HFB energy, the
expressions for these fields are quite simple, whereas with the 
projected-energy
the expressions for these quantities are more involved. 

We consider that the major advantage with the present projection 
formalism is that one can
use the existing HFB computer codes and the only the expressions 
for the fields need
to be redefined. We have applied the projection method to an exactly
soluble deformed single-j shell model with the conclusion that the
numerical effort involed is similar to performing the bare HFB
calculations. The results of the projection method almost agree
completely with the exact shell model calculations and the phase transition
obtained in HFB is smeared out with the projection.

\section{Appendix: Helpful Formulae}

Here, we present some helpful formulae for calculating the derivatives for
matrices and matrix functions which have been extensively used in
the present work. If $A(x)$ and $B(x)$ are arbitrary matrices
depending on a parameter $x$, we have for $A^{\prime }=dA/dx$ 
\begin{eqnarray}
\det (A) &=&e^{{\rm Tr}\ln A}, \\
\frac{d}{dx}\left( AB\right)  &=&A^{\prime }B\,+AB^{\prime }, \\
\frac{d}{dx}A^{-1} &=&-A^{-1}\,A^{\prime }A^{-1}, \\
\frac{d}{dx}\det (A) &=&\det A\,{\rm Tr(}A^{-1}A^{\prime }), \\
\frac{d}{dx}\det (A^{-1}) &=&-\left( \det A^{-1}\right) {\rm Tr(}%
A^{-1}A^{\prime }), \\
\frac{d}{dx}\det (AB) &=&\det (AB)\;{\rm Tr}\left( A^{-1}A^{\prime
}+B^{-1}B^{\prime }\right)  \\
\frac{d}{dx}\det (A^{-1}B) &=&\,\det (A^{-1}B)\;{\rm Tr}\left(
-A^{-1}A^{\prime }+B^{-1}B^{\prime }\right) 
\end{eqnarray}

\newpage

\begin{figure}[t]
%\mbox{\psfig{file=fig1.ps,width=10cm}}
\caption{\label{fig1.fig}
The results of the total energy $(E_{tot})$ and the pair energy 
$(E_{pair})$ for six-particles in a deformed $j=11/2$ orbitial
using different approaches. The projected-HFB results
for $E_{tot}$ are indistinguishable from the exact shell model
calculations.
}

\end{figure}


\newpage

\begin{thebibliography}{9}

\bibitem{thou}
D.J. Thouless,
{\em The Quantum Mechanics of Many-Body Systems\/}
(Academic Press, New York, 1972).

\bibitem{RS.80}
P. Ring and P. Schuck,
{\em The Nuclear Many-Body Problem\/}
(Springer Verlag, New York, 1980).

\bibitem{bethe71}
H.A. Bethe,
Ann. Rev. Nucl. Sci. {\bf 21} (1971) 93.

\bibitem{bcs}
J. Bardeen, L.N. Cooper and J.R. Schrieffer, 
Phys. Rev. {\bf 108} (1957) 5

\bibitem{BMP.58}
A. Bohr, B.R. Mottelson, and D. Pines, Phys. Rev. {\bf 110} (1958) 936

\bibitem{Bel.59}
S.T.Belyaev, Mat. Fys. Medd. {\bf 31} (1959) No.11

\bibitem{bogo}
N.N. Bogoliubov, 
Soviet Phys. JETP {\bf 34} (1958) 41

\bibitem{PY.57} 
R.E. Peierls and J. Yoccoz, Proc. Phys. Soc. {\bf A70} (1957) 381
            
\bibitem{RR.94} 
R. Rossignoli and P.Ring, Ann. Phys. (N.Y.) {\bf 235} (1994) 350

\bibitem{Zeh.67} 
H.D. Zeh, Z. Phys. {\bf 202} (1967) 38

\bibitem{Schmid}  
K.W. Schmid and F. Grummer, Rep. Prog. Phys. {\bf
50} (1987) 731

\bibitem{DMP.64}  
K. Dietrich, H.J. Mang, and J.H. Pradal, Phys. Rev. {\bf
135} (1964) B22

\bibitem{ER.82a} 
J.L. Egido and P.Ring, Nucl.Phys. {\bf A383} (1982) 189,

\bibitem{ER.82b} 
J.L. Egido and P.Ring, Nucl.Phys. {\bf A388} (1982) 19

\bibitem{fo96}
H. Flocard and N. Onishi,
Ann. Phys. (N.Y.) {\bf 254} (1996) 275

\bibitem{val61}
J.G. Valatin,
Phys. Rev. {\bf 122} (1961) 1012

\bibitem{HHR.83}  
A. Hayashi, K. Hara, and P. Ring, Nucl. Phys. {\bf A385 }
(1982) 14

\bibitem{NW.83}  
K. Neerg\u{a}rd and E. W\"{u}st, Nucl. Phys. {\bf A402 }%
(1983) 311

\bibitem{snrp89} 
J.A. Sheikh, M.A. Nagarajan, N. Rowley and K.F. Pal,
Phys. Lett. {\bf B223} (1989) 1

\end{thebibliography}
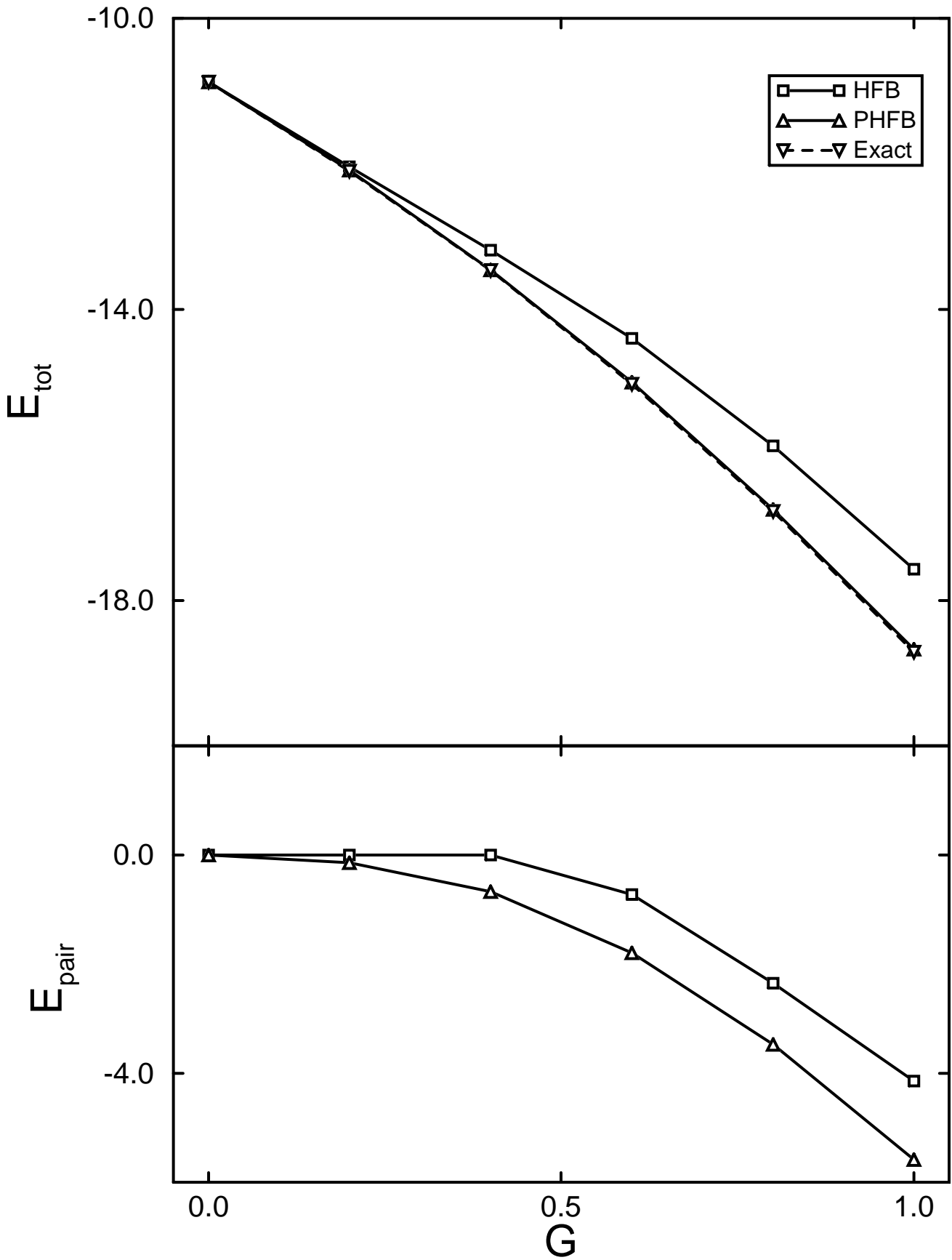
\end{document}